\documentclass[letter]{aa} 

\usepackage{graphicx}
\usepackage{txfonts}
\usepackage{natbib}
\bibpunct{(}{)}{;}{a}{}{,}

\newcommand{\kms}{km\,s$^{-1}$}
\newcommand{\ms}{m\,s$^{-1}$}

\begin{document}

 \title{Carbon $^{12}$C/$^{13}$C isotope ratio of $\alpha$ Aurigae revised}

 \author{Daniel P. Sablowski\thanks{Corresponding author.
     \email{dsablowski@aip.de}}
   \and Silva J\"arvinen
   \and Ilya Ilyin
   \and Klaus G. Strassmeier}

 \institute{Leibniz-Institute for Astrophysics Potsdam (AIP), An der Sternwarte 16, D-14482 Potsdam}

\authorrunning{Sablowski et al.}
\titlerunning{$^{12}$C/$^{13}$C isotope ratio of Capella}

\date{Received 2019; accepted 2019}

 \abstract
   {Capella ($\alpha$~Aur) is one of the few binaries in the sky with two cool giant stars. With spectral types of G8III and G0III, the two components appear at different but distinct stages in their evolution. The G0 secondary star is a Hertzsprung-gap giant, and the G8 primary star is thought to be a clump giant. }
   {We present a new measure of the carbon $^{12}$C/$^{13}$C isotope ratio of the primary component of Capella using high-resolution $R\approx$250\,000 spectra obtained with the Potsdam Echelle Polarimetric and Spectroscopic Instrument (PEPSI) with both the Vatican Advanced Technology Telescope (VATT) and the Large Binocular Telescope (LBT).}
   {Signal-to-noise ratios of up to 2\,700 were obtained by averaging nightly spectra. These average spectra were used to disentangle the two binary components. The isotope ratio was derived with the use of spectrum synthesis from the CN lines at 8004\,\AA .}
   {We found that the $^{12}$C/$^{13}$C ratio of the primary component of Capella is 17.8$\pm$1.9. Our measurement precision is now primarily limited by the spectral-line data and by the grid-step size of the model atmospheres rather than the data. The separated spectrum of the secondary component does not show distinguishable $^{12}$CN and $^{13}$CN lines because of its $v \sin i$ and higher temperature.}
   {Our new $^{12}$C/$^{13}$C value is significantly lower than the previous value of 27$\pm$4 but now agrees better with the recent model prediction of 18.8 - 20.7.}%


\maketitle

\sloppy

\section{Introduction}\label{sec:intro}

Stellar evolution is traced by the carbon, nitrogen, and oxygen (CNO) abundances and in particular by the $^{12}$C/$^{13}$C isotope ratio \citep[e.g.,][]{1538-4357-585-1-L45, doi:10.1093/mnras/stx3027} because carbon will be converted into nitrogen and the $^{12}$C/$^{13}$C ratio will be lowered over time by molecular CN processing (also CH and CO). The surface CNO abundances also depend on the mixing process \citep[e.g.,][]{201218831, 201628138, 201629273}. Moreover, the abundances are not only scaled to stellar nucleosynthesis and mixing, but also to galactic chemical evolution \citep[e.g.,][]{Allende2004}. Previous isotopic measurements \citep[see][]{1981ApJ...248..228L} indicate that the observed ratios for G--K giants are of about 10--30, which is significantly lower than the presumed initial (solar) ratio of about 90  \citep[e.g.][]{ANDERS1989197, Jewitt90}. \cite{1964ApJ...140.1631I} made the first theoretical prediction of evolution-caused changes of isotopic abundances from the reaction
\textsuperscript{12}C($p,\gamma$)\textsuperscript{13}N($\beta^{+}, \nu$)\textsuperscript{13}C($p, \gamma$)\textsuperscript{14}N \citep[see also][]{0004-637X-541-2-660, 0004-637X-846-2-150}. This reaction suggested that \textsuperscript{13}C may be used as an observational tracer once it has been brought up to the surface.
It is therefore expected that the \textsuperscript{12}C/\textsuperscript{13}C ratio decreases and the
\textsuperscript{14}N/\textsuperscript{15}N ratio increases with stellar evolution.

The main mechanism of mixing in stars is convection. However, stars that gain their energy primarily by the CNO cycle have radiative envelopes during their main-sequence lifetime. After hydrogen is exhausted in the core, the core
contracts and heats the outer layers. A convection layer at the surface of the star is established and penetrates the interior of the star. As a consequence, processed material is transported to the surface. This is referred to as the first dredge-up, and it occurs when the star enters the Hertzsprung gap in the Hertzsprung-Russel (HR) diagram. To explain stars with very low carbon isotope ratios, additional mixing processes were suggested. The rotation-induced
meridional mixing can provide mixing even during the main-sequence phase  \citep[e.g.,][]{201628138, Charbonnel2010}. The nature of further mixing processes after the first dredge-up is still debated \citep[e.g.,][]{Pinsonneault1997, Charbonnel2010, 2041-8205-768-1-L11}.

In this respect, binaries are of special interest because they consist of stars of the same age and same initial chemical composition that are at different stages of evolution, if of different mass. Capella (G8III is the primary and G0III is the secondary) is such a system. The combination of the many pieces known for this system, in particular stellar masses at the sub-percent precision \citep{2011A&A...531A..89W}, resulted in rather detailed and precise models of stellar evolution for both components \citep{0004-637X-807-1-26}.
In spite of the small mass difference of only about 3.5$\pm$0.4\% between the two Capella components, they are at very different stages of evolution. The rapidly rotating secondary has just left the main sequence and is on its short path through the Hertzsprung gap, while the primary may be already near the
end of the He-core burning phase. Hence, we expect to see significant differences in the CNO elements and their
isotopes at the surface. The $^{12}$C/$^{13}$C ratio of Capella was measured for the more evolved primary component by \citet{1976ApJ...210..694T} to be 27$\pm$4. This value is now over 40 years old, and its  discrepancy compared to the model values presented by \citet{0004-637X-807-1-26} led to the present study.

We use the CN molecular lines around 8004\,\AA\ for the isotope measurement. This region is the least blended and best accessible wavelength region in the optical for this type of analysis. It displays one line that is purely due to $^{12}$CN and one line that is purely due to $^{13}$CN. New observations are presented in Sect.~\ref{sec:Data}, the data reduction and spectrum disentangling is described in Sect.~\ref{sec:DR}, and our spectrum-synthesis analysis is given in Sect.~\ref{sec:ana}. Our conclusions are summarized in Sect.~\ref{sec:Conclusion}.

\section{Observations}\label{sec:Data}

The observations were carried out with the Potsdam Echelle Polarimetric and Spectroscopic Instrument
\citep[PEPSI;][]{ASNA:ASNA201512172}
installed on the pier of the $2\times 8.4$\,m Large Binocular Telescope (LBT) on Mt.\ Graham, Arizona. Most of the spectra were obtained with the $1.83$\,m Vatican Advanced Technology Telescope (VATT), which has a 450\,m fibre link to the PEPSI spectrograph at the LBT. A Cassegrain adapter on the VATT was used for target acquisition and guiding, and it also provides the feed for the entrance fibre
\citep{SPIE1}.

PEPSI has three resolution modes related to the three fibre core diameters that feed it; see \citet{SPIE2} for a status and most recent mode description of the instrument. We used the ultra-high resolution mode with  $R=\lambda/\Delta\lambda =$250\,000 (2-pixel sampling and 0.74\arcsec\ sky aperture at LBT) with two pairs of $100\,\mu$m fibres and seven-slice image slicers. When fed from the VATT and used before mid-2016, PEPSI provided the same spectral resolution as when fed from the LBT, but with higher losses. After mid-2016, two new pairs of image slicers were implemented for the VATT feed. These provide an expected 2.6-pixel resolution of 200\,000 with a single 3\arcsec\ sky aperture with a $200\,\mu$m fibre and a 9-slice image slicer for a 2-pixel resolution of 250\,000 with the same aperture and a 9-slice image slicer, but smaller slice width. The respective wavelength coverage at LBT and VATT are always identical. PEPSI has blue and red arms with three cross-dispersers (CD) in each arm. One CD in each arm can be used simultaneously.

With the double-eyed LBT the light from the fibre pair of target and sky (optionally, it can be light from a Fabry-P\'erot etalon) is rendered to four image slicers via respective octagonal fibres before it enters the spectrograph. Two 10k$\times$10k STA1600LN CCDs with $9\,\mu$m pixel size and 16 amplifiers optimized for the blue and red arms are employed. When used with the VATT, only a single-eyed light path is available and no sky (but Fabry-P\'erot light). The spectrograph is located in a pressure-controlled chamber at a constant temperature and humidity to keep the refractive index of the air inside constant. This provides a long-term radial velocity stability of about 5\,\ms.

\begin{figure}
    \includegraphics[width=85mm]{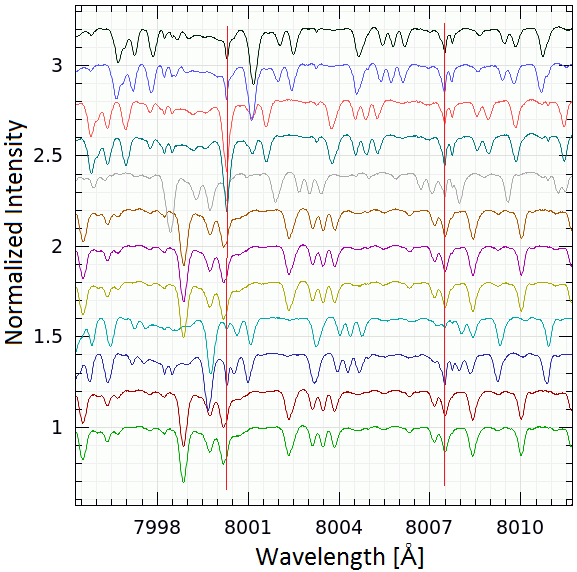}
       \caption{Twelve Capella spectra that were used for the disentanglement. The bottom spectrum is the first spectrum listed in Table~\ref{tab1}. The others are shifted in intensity by +0.2 for better visibility. The spectra are shown in laboratory wavelength scale (prior to heliocentric correction), and some telluric features are marked with  vertical red lines.}
    \label{Fig1}
\end{figure}

We acquired 70 spectra in total of Capella with the red-arm CD\,VI ($\lambda$7355 -- 9137\,\AA) between November 28, 2015 and March 15, 2017. The wavelength solution used about 3\,000 ThAr lines and showed an error of the fit at the image center of 5\,\ms. Consecutive single exposures obtained during the same night were averaged to obtain 12 spectra with high signal-to-noise ratio (S/N)  that were then used to separate the two stellar components. These averaged spectra are listed in Table~\ref{tab1}. The resolving power of the spectra varies according to the setup that was used, between 200k and 250k. The SN/ was measured to be about 8000\,\AA~and is on average about 2\,000:1. Figure~\ref{Fig1} plots the 12 averaged spectra in the 8004\,\AA\ region.

 \begin{table}
   \caption{Overview of observations.}\label{tab1}
  \centering
  \begin{tabular}{lllllll}
  \hline \hline \noalign{\smallskip}
  HJD   & $\phi$& Tel & $R$ & $t_{\text{exp}}$ & S/N & N\\
  (245+)&       &     & $\times10^3$ & (sec)   &     &  \\
  \noalign{\smallskip} \hline \noalign{\smallskip}
  7354.73530 & 0.51 & VATT & 250 & 300 & 2047 & 9\\
  7354.79351 & 0.51 & VATT & 250 & 300 & 1037 & 2\\
  7354.81625 & 0.51 & VATT & 250 & 300 & 2121 & 8\\
  7354.88761 & 0.51 & VATT & 250 & 300 & 2279 & 10\\
  7354.96653 & 0.51 & VATT & 250 & 300 & 745  & 1\\
  7663.00901 & 1.48 & LBT  & 250 & 2.3 & 1740 & 10\\
  7731.95117 & 2.14 & VATT & 200 & 2400& 1377 & 1\\
  7732.97880 & 2.15 & VATT & 200 & 700 & 1354 & 3\\
  7820.72414 & 2.99 & VATT & 250 & 180 & 1616 & 4\\
  7827.69544 & 3.06 & VATT & 250 & 180 & 2245 & 8\\
  8028.00866 & 4.98 & VATT & 250 & 180 & 2638 & 7\\
  8034.94472 & 5.05 & VATT & 250 & 180 & 2738 & 7\\
  \noalign{\smallskip} \hline
  \end{tabular}
  \tablefoot{The S/N is given for the average spectrum from the $N$ single exposures at orbital phase $\phi$ and about 8000\,\AA . The exposure time, $t_{\text{exp}}$, is given in seconds for single exposures.}
 \end{table}

\section{Data reduction and spectral disentangling}\label{sec:DR}%

The data were reduced on a generic software platform written in C++ under Linux, called the Spectroscopic Data System (SDS). The current image processing pipeline is specifically designed to handle the PEPSI data calibration flow and image specific content and is called SDS4PEPSI. The same SDS tools can be used for spectra that are obtained with different \'echelle spectrographs. It provides an automated pipeline without human interaction and relies on statistical inferences and robust statistics. The software numerical toolkit and graphical interface are designed based upon \citet{ilyin2000phd}. Its application to PEPSI has initially been described in \citet{PEPSIdeepII}, and its complete description will be presented in Ilyin (2019, in preparation).

The standard steps of image processing include the following: bias subtraction and variance estimation of the source images, super-master flat-field correction for the CCD spatial noise, \'echelle order definition from the tracing flats, scattered light subtraction, wavelength solution for the ThAr images, optimal extraction of image slicers and cosmic spike elimination of the target image, wavelength calibration and merging slices in each order, normalization to the master flat-field spectrum to remove CCD fringes and blaze function, a global 2D fit to the continuum of the normalized image, and rectification of all spectral orders in the image to a 1D spectrum for a given cross-disperser. The continuum of the final spectra in the time series was corrected for with the use of the mean spectrum. The weighted average of all spectra was normalized to the continuum to eliminate any residual effects in the continuum. The ratio of each individual spectrum and the mean is then used to fit a smoothing spline that constitutes the improved continuum for the individual spectra.

The disentangling code \texttt{Spectangular} was presented by
\citet{refId0}
and is based on singular-value decomposition (SVD) in the wavelength domain. It does not require correct input radial velocities (RVs) because it optimizes them during a full disentangling run. To apply this approach, it is necessary
to use observations that are significantly spread over an orbital period. Because SVD requires an overdetermined set of equations, the minimum input is at least three spectra for two components. The Capella spectra used for the disentangling are shown in Fig.~\ref{Fig1}. Only a small wavelength fraction is plotted. These spectra are not well distributed throughout the orbit ($P_\text{orb}$ = 104~d, see Table \ref{tab1} for covered phases $\phi$), but the small RV shifts between the spectra were sufficient for the algorithm to extract the individual components (see also Appendix \ref{ap2}). The disentangling by the use of poorly sampled orbit was studied in \citet{refId0}. Furthermore, the disentangling code requires a priori knowledge of the light ratio of the two components in the wavelength range under consideration. We adopted the ratio of 1.118 (G8/G0) from \cite{refId10} (see also Appendix~\ref{ap1}).

\begin{figure}
    \includegraphics[width=85mm]{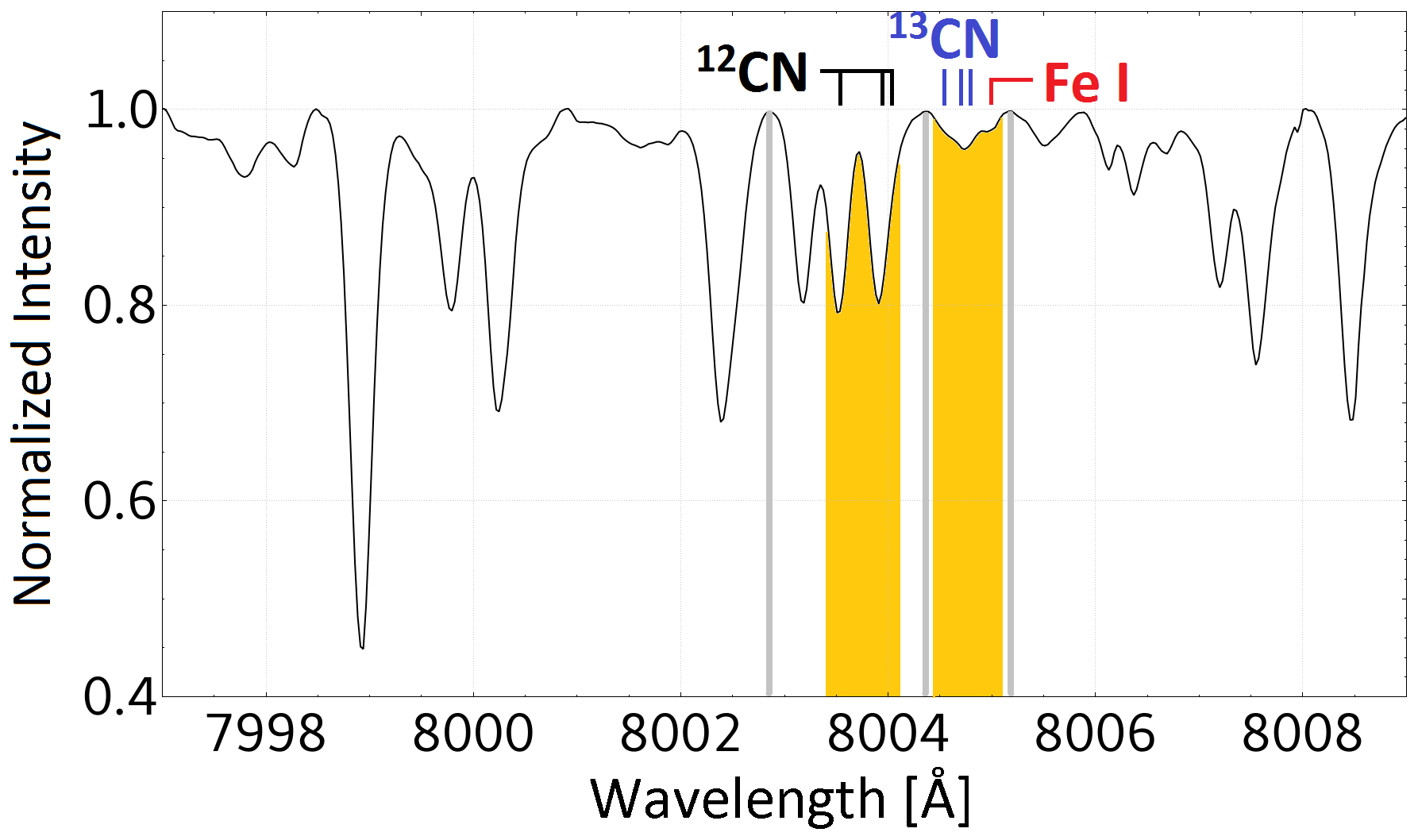}\\
\caption{Separated spectra of the G8III primary for the 8004\,\AA\ wavelength region of Capella. Light gray vertical lines define three continuum points, and the orange regions indicate the two fit regions for the spectrum synthesis.}
    \label{Fig2}
\end{figure}

An additional bonus from this disentangling process is that the telluric spectrum is removed. Water vapor lines, which we call tellurics, further complicate the analysis as well as the disentangling process because they can vary dramatically in strength (and they even slightly vary in RV). This is shown in Fig.~\ref{Fig1} for a pair of telluric lines at wavelengths of about 8007.6\,\AA\ and for a line at 8000.3\,\AA. In order to preserve the flux in the wavelength pixels that are affected by tellurics, we did not apply the heliocentric correction before disentangling. This leads to a fixed position of tellurics in wavelength space. The high variability in the strength of these lines leads  to a strong suppression in the separated spectra, which is due to the least-squares nature of the SVD solution. We then used the residuals from the separated spectra to identify the contribution of tellurics in each spectrum. These residuals were then approximated by a proper spline and corrected in the individual observed spectra. A second run of the entire disentangling process resulted in component spectra with negligible telluric contamination. For more details on this procedure, we refer to a forthcoming paper by \cite{Sablowski19} and to the appendix.
The final result of the separated spectra for the primary of Capella is shown in Fig.~\ref{Fig2}.

\section{Analysis}\label{sec:ana}

\begin{figure}
    \includegraphics[width=85mm]{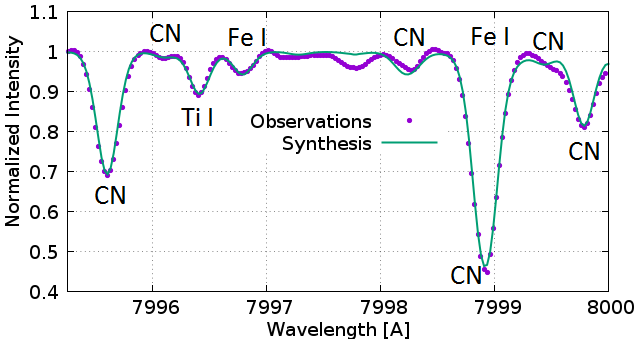}\\
\caption{ Synthesis for refining the values of the broadening mechanisms by a few isolated lines within the separated region. The final values are $\xi_t$ = 1.39 \kms, $\zeta$ = 6.0 \kms, and $v\sin i$ = 3.8 \kms.}
    \label{FigM}
\end{figure}

\begin{figure}
{\bf a.}\\
    \includegraphics[width=85mm]{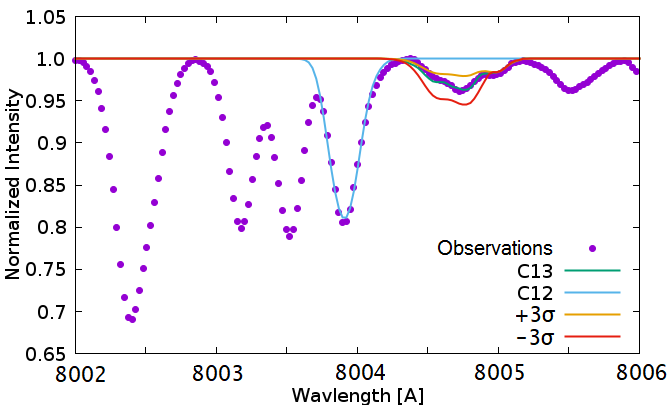}\\
    {\bf b.}\\
    \includegraphics[width=85mm]{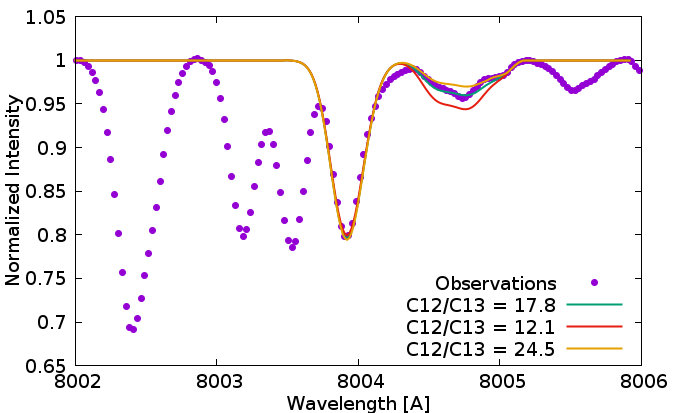}\\
    {\bf c.}\\
    \includegraphics[width=85mm]{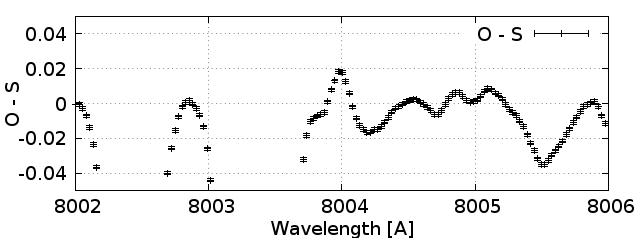}
   \caption{{\bf a:} Result from the spectrum synthesis fitting with \texttt{SME}. The observations are shown as dots, and we plot the best fits as blue and green lines for the $^{12}$CN and $^{13}$CN line region, respectively. This suggests an isotope ratio of 17.8$\pm$1.9. The synthetic spectra for $\pm 3\sigma$ deviations are shown as orange and red lines, respectively.
   {\bf b:} Spectrum synthesis with \texttt{MOOG} for consistency of the spectral fitting by \texttt{SME}. See text for the parameters of the underlying MARCS model.
   {\bf c:} Observations minus synthesis (O-S) for the MOOG synthesis. Error bars corresponds to an S/N of 1,000.}
    \label{Fig3}
\end{figure}

For the carbon isotope ratio (CIR) we employed Spectroscopy Made Easy
\citep[\texttt{SME};][]{1996AAS..118..595V, SME2017}
for a spectrum synthesis under local thermodynamical equilibrium (LTE) together with a grid of spherical model atmospheres from MARCS \citep{refmarcs} for giant stars as provided with the \texttt{SME} code. Because \texttt{SME} is not designed for individual isotope synthesis, we applied the following procedure. First, we fixed the stellar parameters to the values given in Table~\ref{tab2} \citep[as derived by][]{0004-637X-862-1-57, 0004-637X-807-1-26}, but let the carbon abundance remain  a free parameter.
As a result of the high resolution of the spectra, the broadening values had to be reduced slightly compared to those determined by \citet{0004-637X-862-1-57} and \citet{0004-637X-807-1-26}. However, they are within the errors of the given values. These values were obtained by a synthesis of a few isolated lines, see Fig.~\ref{FigM}.
The orbital sampling in other wavelength regions is not sufficient for disentangling. We were therefore unable to determine the atmospheric parameters.
Second, we fit the one line region that is solely due to $^{12}$CN while masking out all the other nearby lines. Third, we fit the other line region that is solely due to $^{13}$CN in the adjacent wavelength bin, again ignoring all the other lines (the fit regions are indicated as shaded areas in Fig.~\ref{Fig2}a). Based on these fits, we found a pseudo- $^{12}$CN and $^{13}$CN abundance from which we obtained the carbon isotope ratio. The resulting best fits for both isotopes are shown in Fig.~\ref{Fig3} along with their 3$\sigma$ deviations.

\begin{table}
  \caption{Stellar parameters for the Capella primary.}\label{tab2}
  \centering
  \begin{tabular}{lllll}
  \hline \hline \noalign{\smallskip}
  Parameter             & Primary & Secondary        & MGS      & Ref.\\
  \noalign{\smallskip} \hline \noalign{\smallskip}
  $T_{\text{eff}}$ (K)  & 4943$\pm$23   & 5694$\pm$73 & 100      & Takeda\\
  $\log g$ (cgs)        & 2.52$\pm$0.08 & 2.88$\pm$0.17& 0.5      & Takeda\\
  $[$Fe/H$]$            & 0.10$\pm$0.05 & -0.08$\pm$0.08& var      & Takeda\\
  $\xi_t$ (\kms)        & 1.47$\pm$0.13 & 2.29$\pm$0.38& 2.0      & Takeda\\
  $\zeta$ (\kms)        & 6.6$\pm$0.4   & (7.0)* & \dots    & Torres\\
  $v\sin i$ (\kms)      & 4.1$\pm$0.5   & 35.0$\pm$0.5& \dots    & Torres\\
  \noalign{\smallskip} \hline
  \end{tabular}
  \tablefoot{Takeda: \citet{0004-637X-862-1-57}.  Torres: \citet{0004-637X-807-1-26}. MGS: model-grid-step size of the used MARCS models. The microturbulence, $\xi_t$ is fixed in the models. The metallicity, $[$Fe/H$]$, has been  varied. *The macroturbulent velocity for the secondary was set to 7.0 \kms.}
\end{table}

We initially implemented the line list from \citet{Carlberg2012} for the \texttt{SME} synthesis. However, another small blend became obvious from our spectra. It is identified with a red thick mark in Fig.~\ref{Fig2}. This line is identified in VALD-3 as \ion{Fe}{i} 8005.0490\,\AA\ \citep{2005MSAIS...8...14K}
with an excitation potential $\chi$=5.58~eV and a transition probability $\log gf=-5.52$. We added it to the \citet{Carlberg2012} list and give the full line list and its parameters in Table~\ref{T3}.

\begin{table}
  \centering
  \caption{Line data for the synthesis.}\label{T3}
  \begin{tabular}{llll}
  \hline \hline \noalign{\smallskip}
  Species & $\lambda$ & $\chi$ & $\log gf$   \\
          &  (\AA)  & (eV)     & (cgs)  \\
  \noalign{\smallskip} \hline \noalign{\smallskip}
  $^{12}$CN & 8003.553 & 0.3109 & -1.6440\\
  $^{12}$CN & 8003.910 & 0.3300 & -1.6478\\
  $^{12}$CN & 8004.036 & 0.0600 & -2.9245\\
  $^{13}$CN & 8004.550 & 0.1200 & -1.5918\\
  $^{13}$CN & 8004.715 & 0.0700 & -2.0814\\
  $^{13}$CN & 8004.801 & 0.1000 & -1.6144\\
  \ion{Fe}{i} & 8005.049 & 5.5869 & -5.518\\
  \ion{Zr}{i} & 8005.248 & 0.6230 & -2.1901\\
 \noalign{\smallskip} \hline
  \end{tabular}
  \end{table}

We note that the precision of our isotope ratio is no longer limited by the data because of its high resolution, high S/N, and a continuum setting that is probably better than 0.1\%. It turned out that the achievable  precision is limited more by the atomic and molecular line data, and probably even more significantly, by the assumptions and grid resolutions of the model atmosphere. Tests were made by measuring the isotope ratio for synthetic spectra from multiple stellar parameters altered by the various grid resolutions and the measurement uncertainties given by
\citet{0004-637X-862-1-57} and \citet{0004-637X-807-1-26}.
\texttt{SME} provides the logarithmic abundances together with their errors, which
for the best fit (stellar parameters from Table~\ref{tab2}) correspond to $\log A_{C_{12}}=-3.18\pm 0.000658$
and $\log A_{C_{13}}=-4.43\pm 0.04598$.

\begin{figure}
    \includegraphics[width=85mm]{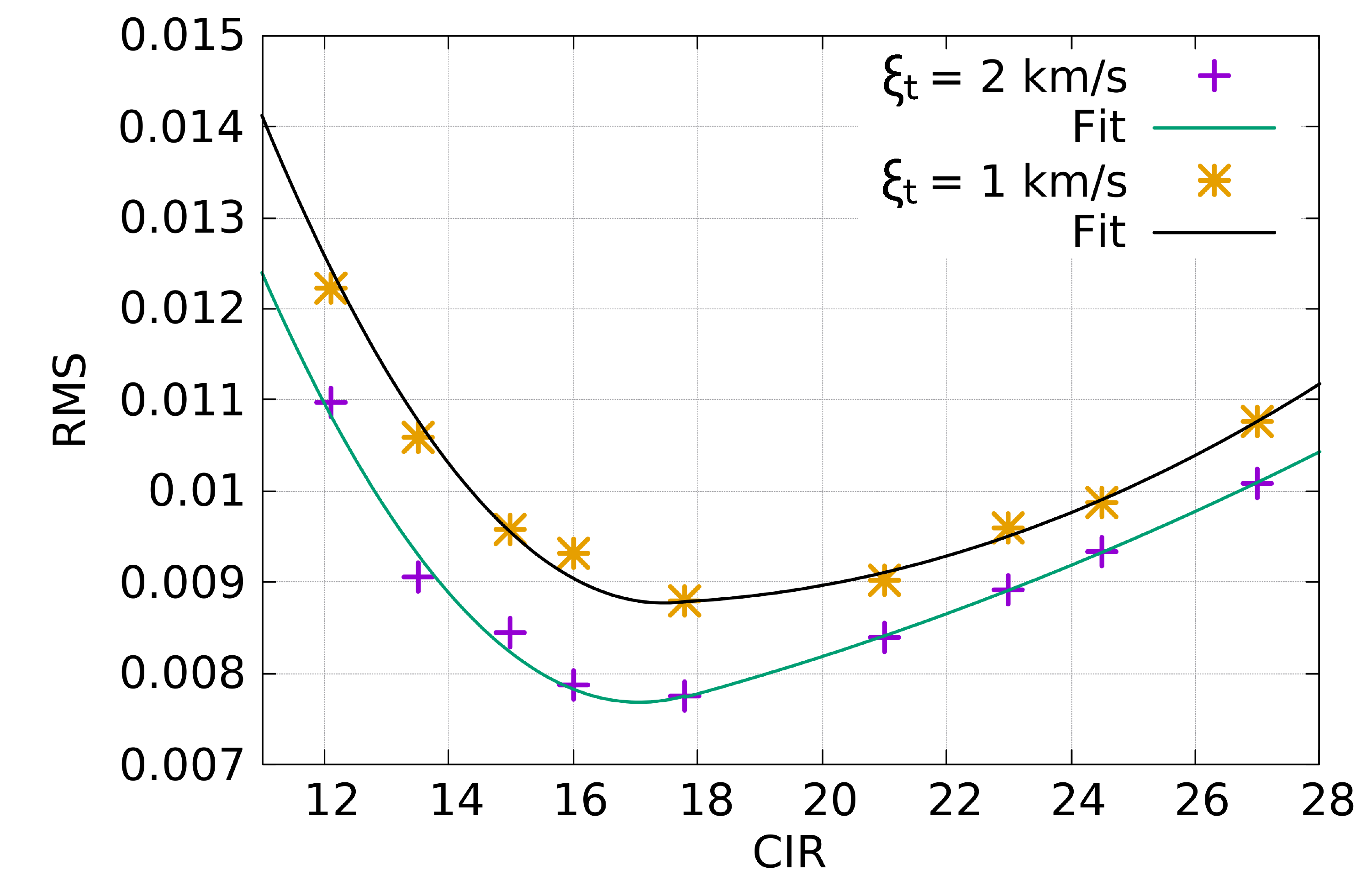}
   \caption{RMS between synthesized spectra from MOOG vs. CIR with the use of MARCS models (see text) with $\xi_t=1.0$~\kms and $\xi_t=2.0$~\kms, respectively.}
    \label{Fig4}
\end{figure}

For a consistency check of the applied method in \texttt{SME}, we generated synthetic spectra with \texttt{MOOG} \citep[][]{Sneden}, which can distinguish between the isotopes. These spectra were created with the same parameters as used in \texttt{SME}. Parameters of the MARCS model selected for the synthesis are $T_{\text{eff}}=5000$\,K, $\log g=2.5$,~$[$Fe/H$]=0.00$,~and $\xi_t=2.0$~\kms. A good agreement is shown by the comparison with the observations in Fig.\,\ref{Fig3}b. The slightly broader lines originate from the higher $\xi_t$ of 2.0 \kms~of this model. Therefore, we also used the model with $\xi_t=1.0$~\kms~and computed synthetic spectra with different carbon isotope ratios (CIR). The  root mean square (RMS) between these spectra and the observations are shown in Fig.\,\ref{Fig4}. The solid line represents a fit of an asymmetric parabola. The minima are in a very good agreement with the CIR value obtained by the fit with \texttt{SME}.

As described above, our analysis is limited to the models and line data. This is shown in Fig.\,\ref{Fig3}c, where we plot the differences between the observations and the synthesis by MOOG with the obtained CIR.

Because the spectra from both codes agree well, we were confident to use the errors provided by the synthesis fitting in \texttt{SME}.
The error of the isotope ratio was calculated according to the error propagation
of the exponential function \citep[e.g.,][]{Bevington} $r=10^{u}$ and $\sigma_r=\ln{10}\cdot r\sigma_u$ , where $r$ is the isotopic ratio, $u=\log A_{C_{12}}-\log A_{C_{13}}$ and
$\sigma^2_u=\sigma^2_{C_{12}}+\sigma^2_{C_{13}}$.
The resulting value is $A_{C_{12}}/A_{C_{13}}=17.8\pm 1.9$.

The separated spectrum of the secondary (see Fig.~\ref{FigB}c) does not show separable signatures of the $^{12}$CN and $^{13}$CN 8004\,\AA\ lines. This is probably due to its higher temperature and much larger $v\sin i$ (see also Appendix~\ref{ap2}).

\section{Summary and conclusion}\label{sec:Conclusion}

We provided new measurement of the carbon $^{12}$C/$^{13}$C isotope ratio of the Capella primary and found it to be 17.8$\pm$1.9. The assigned error is mostly due to uncertainties of the molecular line data and the grid-step-size of the underlying atmospheric models rather than the data. This new ratio is significantly lower than the value of 27$\pm$4 originally measured by \cite{1976ApJ...210..694T} using a curve-of-growth analysis.  \citet{0004-637X-700-2-1349} have pointed out that Tomkin's value was derived with an incorrect light ratio between the two stellar components. They estimated that the equivalent width measurements of the CN lines reported by \cite{1976ApJ...210..694T} were underestimated by about 6--11\% as a result of this effect. Because the equivalent widths were used strictly differentially, we second the statement of  \citet{0004-637X-700-2-1349} that it introduced only a small error that is likely within the error bars already given. The difference to our new ratio therefore likely originates from the improvement of the data, the synthesis analysis, and better spectral-line data.

Although our new isotope ratio is now in a better agreement with theoretical expectations of 20.7 by \citet{0004-637X-807-1-26} based on MESA stellar evolution models, it is now (marginally) below the predictions if at an age of 588.5~Myr \citep[see][]{0067-0049-208-1-4}. The earlier prediction by \citet{0004-637X-700-2-1349} was based on models from \citet{Claret2004} and \citet{0004-637X-618-2-908}, which were all in the range of 18.88 to 19.60.

\begin{acknowledgements}
We thank the anonymous referee for the constructive comments that helped to improve the content of this study.
We also thank our colleagues Gohar Harutyunyan, Arto J\"arvinen, and the developer of \texttt{MOOG}, Christopher A. Sneden, as well as the developers of \texttt{SME}, Nikolai Piskunov and Jeff Valenti, for their kind support to get the synthesis codes \texttt{MOOG} and \texttt{SME} running.
This work is based on data acquired with PEPSI at the Large Binocular Telescope (LBT)
  and with the Vatican Advanced Technology Telescope (VATT). The LBT is an
  international collaboration among institutions in the United States, Italy,
  and Germany. LBT Corporation partners are the University of Arizona on
  behalf of the Arizona university system; Istituto Nazionale di Astrofisica,
  Italy; LBT Beteiligungsgesellschaft, Germany, representing the Max-Planck
  Society, the Leibniz-Institute for Astrophysics Potsdam (AIP), and
  Heidelberg University; the Ohio State University; and the Research
  Corporation, on behalf of the University of Notre Dame, University of
  Minnesota, and University of Virginia. This work has made use of the VALD database, operated at Uppsala University, the Institute of Astronomy RAS in Moscow, and the University of Vienna.
\end{acknowledgements}

\bibliographystyle{aa}
\bibliography{my-library}

\begin{appendix}
 \section{Correction of tellurics and flux ratio}\label{ap1}

 Because the correction of the telluric lines in the wavelength range under consideration is crucial for the analysis, some details of this procedure are described here.
 A forthcoming paper by \cite{Sablowski19} will discuss these disentangling problems in more detail.

\begin{figure}
{\bf a.}\\
    \includegraphics[width=85mm]{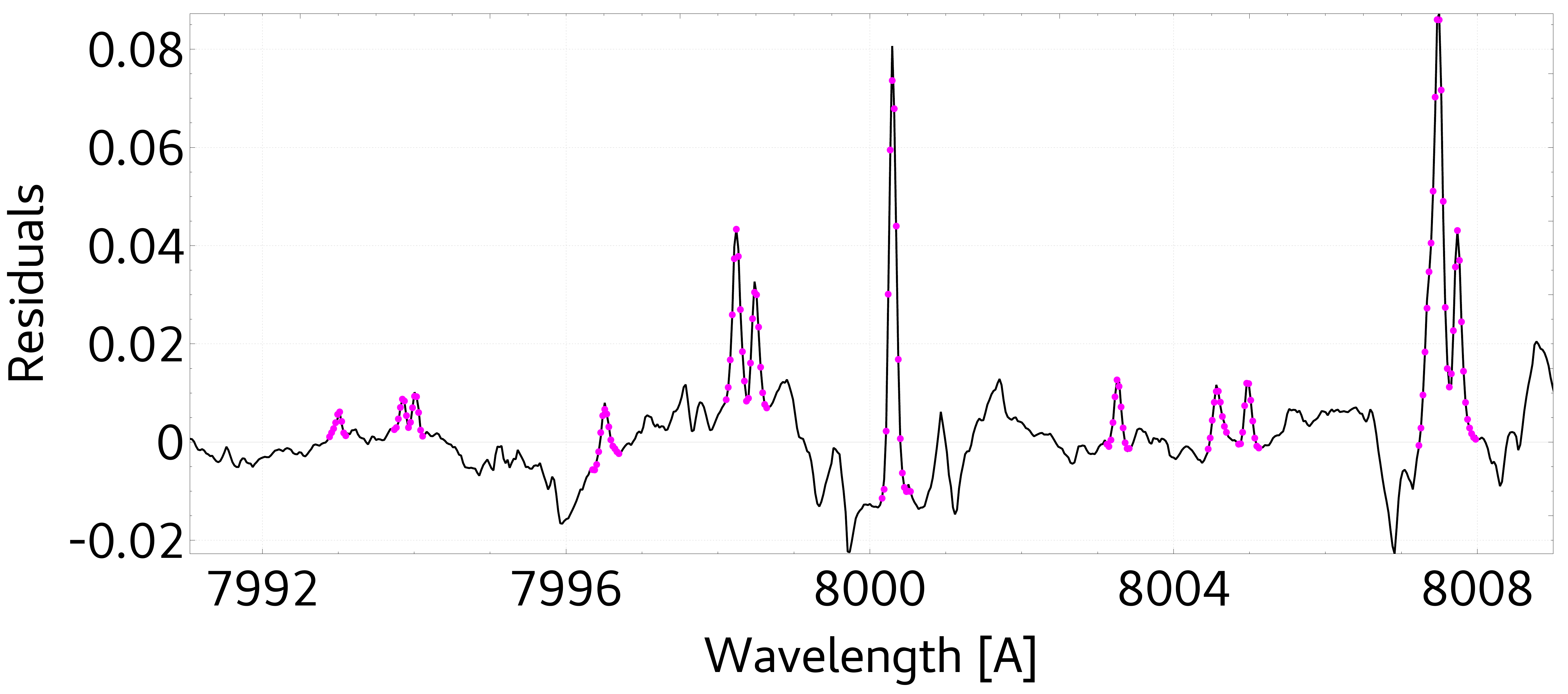}

    {\bf b.}\\
    \includegraphics[width=85mm]{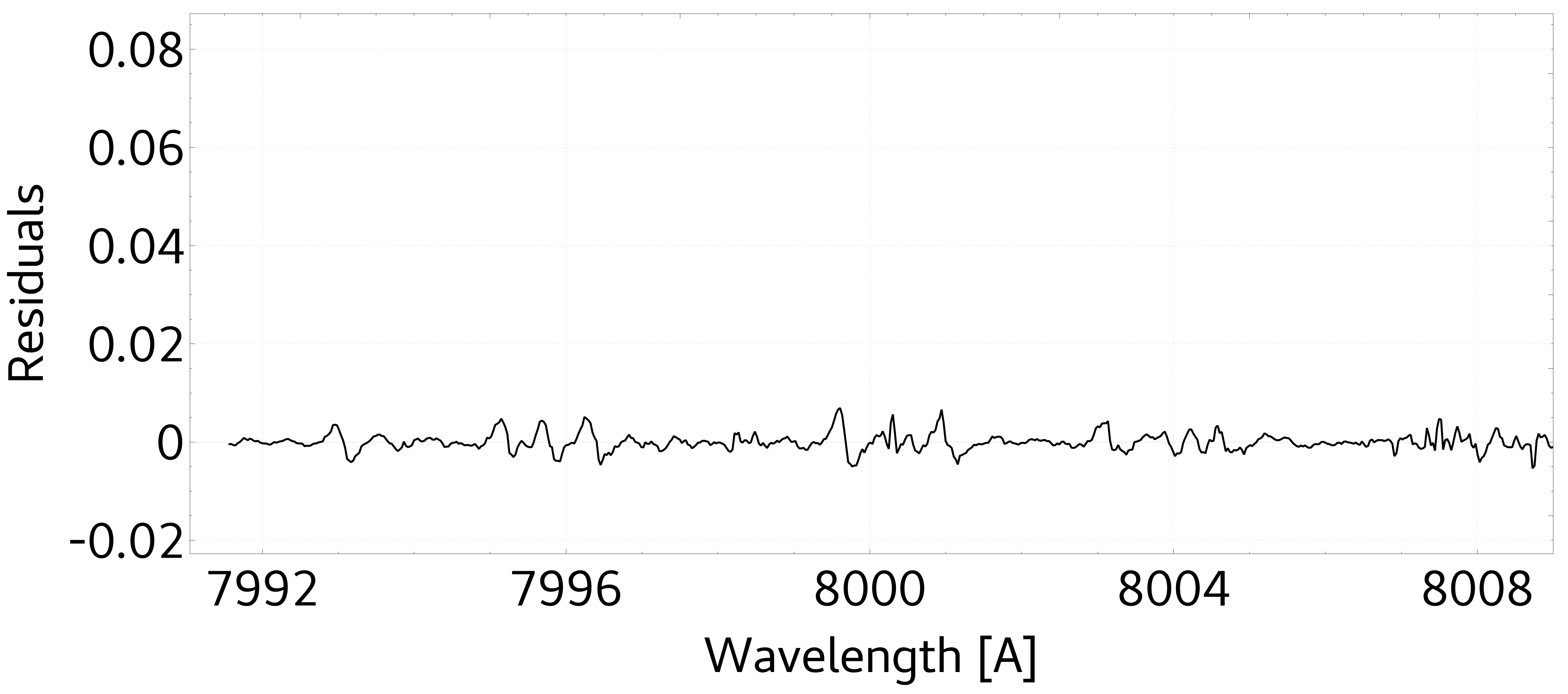}
   \caption{Residual spectrum between separated spectra of the two components and an observation with strong telluric lines. {\bf a.} Residuals before correction. The dots represent the spline used to approximate the telluric lines. {\bf b.} Residuals to the final corrected spectrum.}
    \label{FigA1}
\end{figure}

 The residual spectrum between the separated spectra and an observation is shown in Fig.~\ref{FigA1}a. The identified tellurics are approximated by spline fits shown by dots. These splines were used to correct for the telluric lines in the observations. Special care was taken to identify the telluric lines such that no stellar signatures were affected. 
 This was ensured by an iterative process: Only one line (we took the strongest) at a time was removed. The disentangling was performed again, and the next line with the highest contribution to the residuals was corrected. 
 In this way, we corrected for all the telluric lines indicated by the dots in Fig.~\ref{FigA1}a in all 12 observed spectra. The same residuals for the fully corrected observations after the final disentangling run is shown in Fig.~\ref{FigA1}b.

  \begin{table}
  \small
   \caption{Relative light ratio measurements for Capella.}
   \begin{tabular}{llll}
   \hline \noalign{\smallskip}
      \hline \noalign{\smallskip}
  Reference & \multicolumn{2}{c}{Wavelength [\AA]} & $k=$  \\
   ~ & start & end & $l_{G8}/l_{G0}$\\
  \noalign{\smallskip} \hline \noalign{\smallskip}
  \cite{1988AJ.....96.1056B} & 3996.5 & 4219.5 & 0.61 $\pm$ 0.02\\
  \cite{1986JApA....7...45G} & 3931 & 4919 & 0.69  \\
  \cite{1994AJ....107.1859H} & 4400 & 4600 & 0.77 $\pm$ 0.07\\
  \cite{1988AJ.....96.1056B} & 4568 & 4770 & 0.82 $\pm$ 0.02\\
  \cite{0004-637X-700-2-1349} & 5165.5 & 5210.5 & 0.677 $\pm$ 0.023\\
  \cite{1983LowOB...9..185H} & 5150 & 5250 & 0.64 $\pm$ 0.03\\
  \cite{0004-637X-700-2-1349} & 4086 & 6464 & 0.854 $\pm$ 0.041\\
  \cite{1988AJ.....96.1056B} & 5351.5 & 5604.5 & 0.93 $\pm$ 0.02\\
  \cite{1988AJ.....96.1056B} & 5351.5 & 5604.5 & 0.87 $\pm$ 0.02\\
  \cite{1988AJ.....96.1056B} & 5351.5 & 5604.5 & 0.92 $\pm$ 0.02\\
  \cite{1988AJ.....96.1056B} & 5351.5 & 5604.5 & 0.91 $\pm$ 0.02\\
  \cite{1994AJ....107.1859H} & 5383 & 5583 & 0.87 $\pm$ 0.04\\
  \cite{1983LowOB...9..185H} & 5450 & 5550 & 0.82 $\pm$ 0.02\\
  \cite{1983LowOB...9..185H} & 5420 & 5620 & 0.89 $\pm$ 0.01\\
  \cite{1986JApA....7...45G} & 5095 & 5993 & 0.87 \\
  \cite{refId10} & 6112 & 8430 & $\lambda$-dependent \\
  \cite{1994AJ....107.1859H} & 8003 & 8203 & 1.05 $\pm$ 0.05\\
  \noalign{\smallskip} \hline
   \end{tabular}
   \label{TA1}
  \end{table}

\begin{figure}
    \includegraphics[width=85mm]{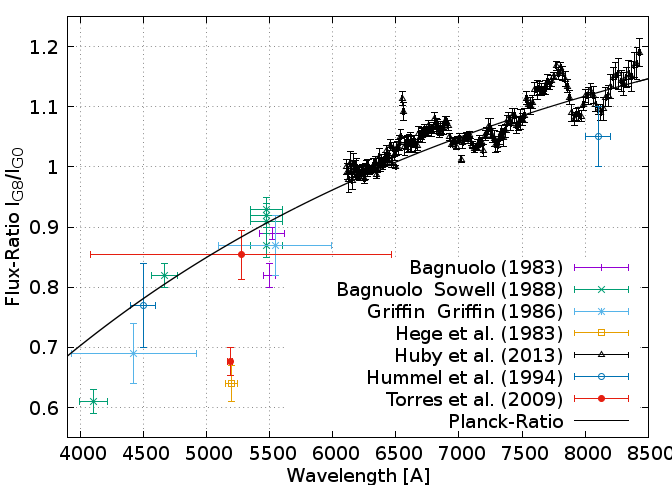}
   \caption{Visual presentation of the measurements of the light ratio for Capella (see also Table~\ref{TA1}).}
    \label{FigA2}
\end{figure}

Measurements of the light ratio for Capella are summarized in Table~\ref{TA1} and plotted in Fig.~\ref{FigA2}. The black solid line is the ratio $r$ given by dividing two Planck relations $r = aP(T_{\text{eff,p}})/P(T_{\text{eff,s}})$ with temperatures as given in Table~\ref{tab2}.
A multiplication factor $a = 1.04$ is needed to adapt to the data.

\section{Spectrum of the secondary}\label{ap2}

\begin{figure}
{\bf a.}\\
\includegraphics[width=85mm]{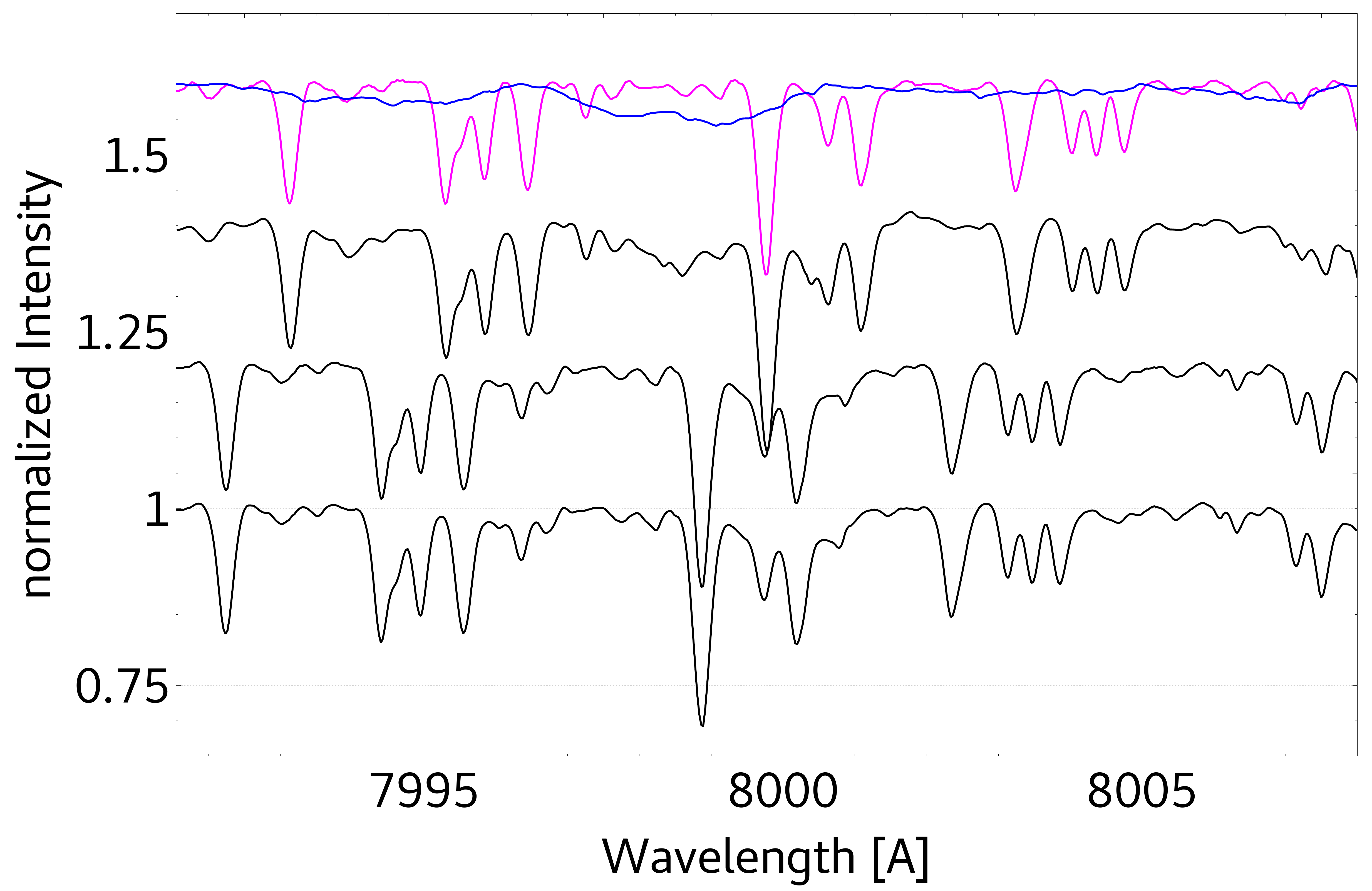}\\
{\bf b.}\\
\includegraphics[width=85mm]{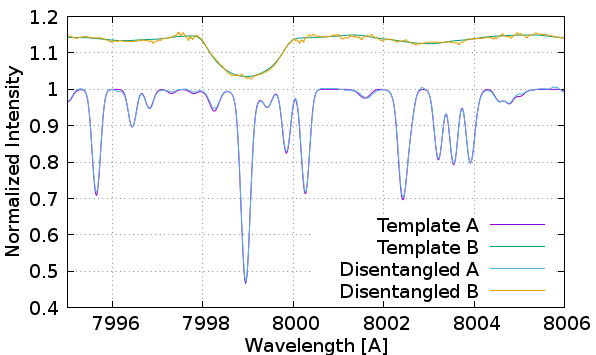}\\
{\bf c.}\\
    \includegraphics[width=85mm]{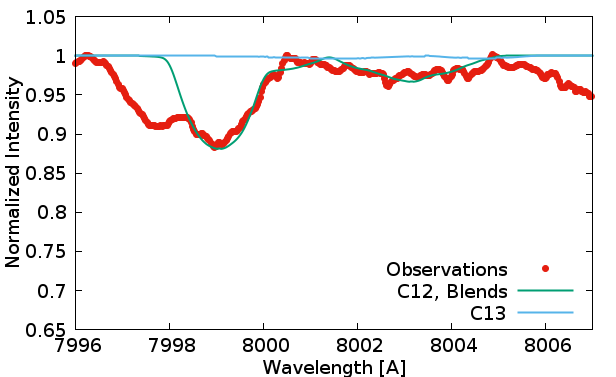}
       \caption{{\bf a:} Three selected observations (each shifted on y by 0.2 for better visibility) with the weakest telluric feature (black) and the separated spectra (purple and blue) from these observations. The separated spectra are shifted in wavelength to match the topmost observation. The separated spectra are scaled such that the line strength matches the strengths in the observations. {\bf b:} Synthetic spectra (used as templates to create an artifical binary data set) in comparison with the separated spectra (see Text). {\bf c:} The separated spectrum (red) of the G0III secondary of Capella with a synthetic spectrum (green). This was derived with the stellar parameters as listed in Table~\ref{tab2} and the line list given in Table~\ref{TB1}. It shows predominantly only the strong \ion{Fe}{i} 7999\,\AA\,line. The blue line is a synthetic spectrum based on the $^{13}$CN lines as with an abundance as derived for the primary.}
    \label{FigB}
\end{figure}

To ensure that the disentangling is not affected by the tellurics or by the limited sampling of the orbit, further tests were performed. First, three spectra that show the smallest signal from tellurics were selected from the 12 spectra listed in Tabel \ref{tab1}. As pointed out by \citet{refId0}, at least 3 spectra are required because SVD has to be performed on an overdetermined system of equations, that is, for two components present in the spectra, at least three observations are required. Furthermore, these spectra should correspond to distinctive orbital phases. The separated spectra obtained from that data set are in excellent agreement with the spectra used for the synthesis (see Fig.\,\ref{FigB}a).
Second, to show that the orbital sampling by the observations is sufficient for the disentangling, 12 spectra were created by the use of synthesized spectra covering the same orbital phases as the observed spectra listed in Table \ref{tab1}. The excellent agreement between the templates (the synthesized spectra) and the separated spectra is shown in Fig.\,\ref{FigB}b.

The weak feature around the 8004\,\AA\ lines can almost entirely be reproduced by the blend lines (\ion{Fe}{i} 8002.6\,\AA\,, 8003.2\,\AA\,, \ion{Al}{i} 8003.2\,\AA,\, and \ion{Ti}{i} 8003.5\,\AA) 
and with a carbon abundance as derived for the primary.
However, we show the separated spectrum together with a theoretical spectrum in Fig.~\ref{FigB}c. We used the parameters given in Table~\ref{tab2} and a macroturbulent velocity of 7.0 \kms.
The used line list is a combination of the list provided by \citet{Carlberg2012} and selected lines (with significant strengths) from VALD-3. The blue-side line to the strong \ion{Fe}{i} 7998.944\,\AA\,
could not be identified. It is too strong for a $^{12}$CN line (it would disagree with lines at 8004\,\AA). VALD-3 lists a central depth of the fitted \ion{Fe}{i} 7998.944 of 0.534. There are only two lines listed in VALD-3 with significant depth: \ion{Fe}{i} 7996.8156\,\AA\,with a depth of 0.026 and \ion{Ti}{i} 7996.4350\,\AA\,with a depth of 0.072, but they are too weak to fit the observation.
The NIST atomic line data base \citep{NIST_ASD} lists only one further and rather exotic posibility of an \ion{Th}{i} line with unknown $\log gf$ value. A synthesis would lead to unrealistic abundance and high $\log gf$ value for that line.
Furthermore, we plot the synthetic spectrum of the $^{13}$CN lines (blue line, abundance as derived for the primary).

\begin{table}
\small
  \centering
  \caption{Line data for the synthesis of the secondary. Compiled with data from \citet{Carlberg2012} and VALD-3.}\label{TB1}
  \begin{tabular}{llll}
  \hline \hline \noalign{\smallskip}
  Species & $\lambda$ & $\chi$ & $\log gf$   \\
          &  (\AA)  & (eV)     & (cgs)  \\
  \noalign{\smallskip} \hline \noalign{\smallskip}
  $^{12}$CN & 7990.388 & 1.380 & -2.0585\\
  $^{12}$CN & 7990.790 & 1.450 & -1.6234\\
  $^{12}$CN & 7991.583 & 1.470 & -1.6216\\
  $^{12}$CN & 7992.297 & 0.090 & -2.0114\\
  $^{12}$CN & 7992.297 & 0.180 & -1.5143\\
  \ion{Ti}{i} & 7993.600 & 1.873 & -2.4970\\
  $^{12}$CN & 7994.018 & 1.420 & -1.6326\\
  \ion{Fe}{i} & 7994.459 & 5.946 & -2.242\\
  $^{12}$CN & 7994.694 & 0.110 & -1.9666\\
  \ion{Ni}{i} & 7994.775 & 5.481 & -1.175\\
  $^{12}$CN & 7995.640 & 0.060 & -2.9172\\
  $^{12}$CN & 7995.640 & 0.290 & -1.6556\\
  $^{12}$CN & 7995.640 & 0.310 & -1.6615\\
  \ion{Ti}{i} & 7996.435 & 3.337 & 0.2660\\
  \ion{Fe}{i} & 7996.8156 & 4.5844 & -2.505\\
  \ion{Sc}{i} & 7997.372 & 5.900 & -5.489\\
  \ion{Ti}{i} & 7997.481 & 4.515 & -2.311\\
  $^{12}$CN & 7998.312 & 1.540 & -1.2480\\
  \ion{Fe}{i} & 7998.944 & 4.371 & 0.1489\\
  $^{12}$CN & 7999.214 & 1.400 & -2.0287\\
  $^{12}$CN & 7999.214 & 1.600 & -1.8041\\
  $^{12}$CN & 7999.846 & 0.100 & -1.9830\\
  $^{12}$CN & 8000.261 & 0.190 & -1.4962\\
  $^{12}$CN & 8000.316 & 1.470 & -1.6091\\
  \ion{Fe}{i} & 8000.317 & 5.538 & -2.878\\
  \ion{Fe}{i} & 8000.396 & 6.038 & -1.194\\
  \ion{Nd}{ii} & 8000.757 & 1.091 & -1.2220\\
  $^{12}$CN & 8001.524 & 1.420 & -1.6253\\
  $^{12}$CN & 8001.652 & 1.480 & -1.6091\\
  $^{12}$CN & 8002.412 & 0.180 & -1.4962\\
  \ion{Fe}{i} & 8002.576 & 4.580 & -2.2400\\
  \ion{Al}{i} & 8003.185 & 4.087 & -1.8790\\
  $^{12}$CN & 8003.213 & 0.120 & -1.9431\\
  \ion{Fe}{i} & 8003.226 & 5.539 & -1.7630\\
  \ion{Ti}{i} & 8003.485 & 3.724 & -0.2000\\
  $^{12}$CN & 8003.553 & 0.311 & -1.6440\\
  $^{12}$CN & 8003.213 & 0.120 & -1.9431\\
  $^{12}$CN & 8003.910 & 0.3300 & -1.6478\\
  $^{12}$CN & 8004.036 & 0.0600 & -2.9245\\
  \ion{Fe}{i} & 8005.049 & 5.5869 & -5.518\\
  \ion{Zr}{i} & 8005.248 & 0.623 & -2.1901\\
  \ion{Si}{i} & 8006.459 & 6.261 & -1.7220\\
  \ion{Fe}{i} & 8006.702 & 5.067 & -2.2820\\
  $^{12}$CN & 8006.925 & 1.600 & -1.7878\\
  \ion{Co}{i} & 8007.242 & 4.146& -0.1159\\
 \noalign{\smallskip} \hline
  \end{tabular}
  \end{table}

\end{appendix}

\end{document}